\documentclass[11pt]{article}

\usepackage[a4paper,margin=28mm,headheight=14pt]{geometry}
\usepackage[T1]{fontenc}
\usepackage{lmodern}
\usepackage{amsmath,amssymb,amsthm,mathtools}
\usepackage{aliascnt}
\usepackage{needspace}
\usepackage{microtype}
\usepackage[authoryear,round]{natbib}
\usepackage{xcolor}
\usepackage{fancyhdr}
\usepackage[colorlinks=true,linkcolor=blue!45!black,
  citecolor=blue!45!black,urlcolor=blue!45!black]{hyperref}
\usepackage[nameinlink,noabbrev]{cleveref}

\newtheorem{theorem}{Theorem}[section]
\newaliascnt{proposition}{theorem}
\newtheorem{proposition}[proposition]{Proposition}
\aliascntresetthe{proposition}
\newaliascnt{lemma}{theorem}
\newtheorem{lemma}[lemma]{Lemma}
\aliascntresetthe{lemma}
\newaliascnt{corollary}{theorem}

\aliascntresetthe{corollary}
\theoremstyle{definition}
\newaliascnt{remark}{theorem}

\aliascntresetthe{remark}
\crefname{theorem}{Theorem}{Theorems}
\crefname{proposition}{Proposition}{Propositions}
\crefname{lemma}{Lemma}{Lemmas}
\crefname{corollary}{Corollary}{Corollaries}
\crefname{remark}{Remark}{Remarks}

\newcommand{\R}{\mathbb R}
\newcommand{\C}{\mathbb C}
\newcommand{\ii}{\mathrm i}
\newcommand{\dd}{\,\mathrm d}
\newcommand{\Hess}{\operatorname{Hess}}
\newcommand{\cof}{\operatorname{cof}}
\newcommand{\ord}{\operatorname{ord}}
\newcommand{\Crit}{\operatorname{Crit}}
\newcommand{\cQ}{\mathcal Q}
\newcommand{\cX}{\mathcal X}
\newcommand{\cZ}{\mathcal Z}

\hypersetup{
  pdftitle={Equilibrium finiteness and Saari's conjecture for the planar circular restricted N-body problem},
  pdfauthor={},
  pdfsubject={Unconditional planar equilibrium finiteness and constant-amended-potential rigidity}}

\title{Equilibrium finiteness and Saari's conjecture\protect\\
for the planar circular restricted \(N\)-body problem}
\author{Qinglong Zhou$^{1,2}$\thanks{Partially supported by NSFC (No.12171426), the Natural Science Foundation of Zhejiang Province (No. LY19A010020) and the Fundamental Research Funds for the Central Universities (No. 226-2024-00136).
		E-mail: zhouqinglong@zju.edu.cn. }\\
	$^{1}$ School of Mathematical Sciences,\\Zhejiang University, Hangzhou 310058, Zhejiang, China\\
    $^{2}$ Institut f\"ur Mathematics,\\Augsburg Universit\"at, 86159 Augsburg, Germany\\
}
\date{}

\begin{document}
\maketitle
\vspace{-2.5ex}

\begin{abstract}
We establish two results for planar Newtonian centers in uniform
rotation. First, the associated quadratic-plus-Newtonian amended
potential has finitely many equilibria for every finite configuration
of at least two distinct centers with positive masses. 
Second, every
collision-free planar test-particle motion with constant amended
potential, equivalently constant rotating-frame speed, is stationary
in the rotating frame.
When the centers are primary bodies in a circular relative equilibrium,
this proves the amended-potential form of Saari's conjecture for the
restricted \(N\)-body problem. 
\end{abstract}



\section{Introduction}

This paper concerns two questions for planar quadratic-plus-Newtonian
potentials: whether their equilibrium sets are finite for every choice
of distinct centers, and whether a test particle in a uniformly
rotating frame can move nontrivially along a potential level. The
second question is the restricted version of Saari's conjecture. We
first state the equilibrium-finiteness theorem and then formulate the
restricted version of Saari's conjecture together with its celestial-mechanical background.

\citet{KulevichRobertsSmith2009} proved finiteness of equilibria for the
triangular planar circular restricted four-body problem. More generally,
\citet[Theorem~3.1]{HamptonJensen2015} proved finiteness of
central-configuration extensions of a fixed planar subconfiguration,
explicitly allowing the added body to have zero mass. Their theorem
therefore already implies finiteness of the real restricted equilibria
for fixed central-configuration primaries. We regard these as known
results. Finiteness of equilibria and Saari rigidity are nevertheless
distinct: a finite critical set does not by itself exclude a
nonstationary orbit on a regular potential level.

The potential considered throughout the paper is
\begin{equation}\label{eq:Omega-intro}
 \Omega(q)=\frac{\omega^2}{2}|q|^2+
                   \sum_{j=1}^n\frac{m_j}{|q-a_j|},
 \qquad q\in\R^2\setminus\{a_1,\ldots,a_n\},
\end{equation}
where the centers \(a_j\) are any pairwise distinct real planar points
and \(\omega\in\R\setminus\{0\}\). The primary central-configuration
equations are needed to obtain a Newtonian restricted problem, but they
are not used in the finiteness argument. Our first main result concerns
the critical points of \eqref{eq:Omega-intro} for arbitrary centers.

\begin{theorem}\label{thm:equilibria}
For \(n\ge2\), suppose \(a_1,\ldots,a_n\in\R^2\) are pairwise distinct
and not necessarily in a central configuration.
Let \(m_1,\ldots,m_n>0\), and
\(\omega\in\R\setminus\{0\}\). Then the potential
\eqref{eq:Omega-intro} has finitely many critical points in
\(\R^2\setminus\{a_1,\ldots,a_n\}\).
\end{theorem}
\citet[Theorem~2.1]{ArustamyanEtAl2021} proved that the critical points
are finite and nondegenerate outside a measure-zero set of parameters.
They explicitly left open whether finiteness holds for every
configuration with \(n\ge2\); see their Section~2.1 and concluding
remarks. That work already allows arbitrary planar center locations:
relative to it, the advance is not the introduction of arbitrary
centers, but removal of the exceptional parameter set. Their
normalization \(\omega^2=1\) is equivalent to ours for this question:
dividing \(\Omega\) by \(\omega^2>0\) replaces \(m_j\) by
\(m_j/\omega^2\) without changing the critical set.
\Cref{thm:equilibria} settles the stated qualitative finiteness question
for distinct positive planar centers, including exceptional parameter
values with degenerate equilibria. It does not assert that every
equilibrium is nondegenerate.

\citet[Theorem~4]{Zolotov2023} subsequently obtained an upper bound on
the number of isolated critical points of the same type of potential
in \(\R^d\), using the Thom--Milnor theorem. An upper bound for
isolated critical points does not exclude a positive-dimensional
critical set, an issue also discussed in his Section~4.4.
Together, these two results for arbitrary center configurations provide
generic finiteness and bounds for isolated equilibria, but neither
excludes a critical curve at exceptional parameter values in that
class. The contribution of \cref{thm:equilibria}
is precisely this exclusion for every admissible planar configuration.
The result complements those quantitative estimates; it is not an
improved numerical bound. Our argument proves more: the entire complex
critical locus on the collision-free distance cover is finite, as
stated in \cref{thm:complex-equilibria}. In particular, the obstruction
is algebraic and does not depend on restricting a hypothetical curve
to the real physical sheet.

The distinction from \citet{HamptonJensen2015} is also essential.
When the primaries form a central configuration, their zero-mass
extension theorem already gives real finiteness. In
\cref{thm:equilibria}, however, the centers are arbitrary and need not
satisfy any of the massive-body central-configuration equations.
The three comparisons identify different aspects of the contribution:
arbitrary centers extend the central-configuration setting of
\citet{HamptonJensen2015}; unconditional finiteness removes the
genericity restriction of \citet{ArustamyanEtAl2021}; and exclusion of
nonisolated sets supplies information not furnished by the isolated-point
bounds of \citet{Zolotov2023}. The principal advance is therefore
unconditional qualitative finiteness in the full planar class, with
complex critical-locus finiteness providing a stronger algebraic
statement.

We next turn to Saari's conjecture for the
Newtonian $N$-body problem which asserts that a solution with constant
moment of inertia about its center of mass is a relative equilibrium
\citep{Saari1970}. The Lagrange--Jacobi identity forces the Newtonian
potential, and hence the kinetic energy, to be constant along such a
solution. This observation leads to an algebraic approach: derive
conditions on the mutual distances and exclude positive-dimensional
families satisfying all of them. This is the approach used by
\citet{Moeckel2005} in the planar three-body problem.

A restricted problem requires a different formulation. An infinitesimal
particle does not contribute to the total moment of inertia, so
constancy of that inertia cannot constrain its trajectory once the
primaries have been prescribed. \citet{RobertsMelanson2007} formulated
the restricted analogue in terms of the amended potential in a
uniformly rotating frame. They proved that a solution of the planar
circular restricted three-body problem cannot travel along a level
curve of the amended potential unless it is an equilibrium. By the
Jacobi integral, this is equivalent to excluding nonstationary solutions
of constant speed in that frame. Their proof eliminates the velocity
and applies Bernstein--Khovanskii--Kushnirenko theory to polynomial
equations in the two primary distances. A restricted four-body
extension was obtained in the thesis of \citet{Bruder2008}, as recorded
by \citet{KulevichRobertsSmith2009}.

Our second main result extends this amended-potential rigidity to any
number of primaries in a planar central configuration, realized as a
circular relative equilibrium. For its statement, let \(R(\theta)\)
denote counterclockwise planar rotation through angle \(\theta\), and
let \(J(x,y)=(-y,x)\). A central configuration in center-of-mass
coordinates satisfies
\[
 \sum_{j=1}^n m_ja_j=0,\qquad
 \sum_{\ell\ne j}m_\ell
 \frac{a_\ell-a_j}{|a_\ell-a_j|^3}=-\lambda a_j
 \quad(j=1,\ldots,n),
\]
where \(\lambda>0\). Choosing \(\omega\in\R\setminus\{0\}\) with
\(\omega^2=\lambda\) gives the circular primary motion
\(Q_j(t)=R(\omega t)a_j\).

\Needspace{20\baselineskip}
\begin{theorem}[Restricted Saari theorem]\label{thm:restricted}
Let \(n=N-1\ge2\), and let \(n\) positive-mass primaries form a
collision-free planar central configuration in center-of-mass
coordinates. Realize it as the circular relative equilibrium
\(Q_j(t)=R(\omega t)a_j\), where \(\omega^2=\lambda\).
Let \(\mathcal T\subset\R\) be a connected open interval, and let
\(q:\mathcal T\to\R^2\setminus\{a_1,\ldots,a_n\}\) be a \(C^2\)
solution of
\begin{equation}\label{eq:rotating-intro}
 \ddot q+2\omega J\dot q=\nabla\Omega(q),
\end{equation}
with \(\Omega\) given by \eqref{eq:Omega-intro}. Then the following
are equivalent:
\begin{enumerate}
 \item the amended potential \(\Omega(q(t))\) is constant;
 \item the rotating-frame speed \(|\dot q(t)|\) is constant;
 \item \(q(t)\equiv q_*\) for an equilibrium
       \(\nabla\Omega(q_*)=0\).
\end{enumerate}
In inertial coordinates, such a solution is
\(Q(t)=R(\omega t)q_*\), so it rotates rigidly with the primaries.
\end{theorem}

Theorem~\ref{thm:restricted} concerns planar motion and a circular
realization of the primary central configuration. No symmetry,
nondegeneracy, or restriction on the Jacobi integral is assumed. It
does not assert an elliptic-primary or spatial version of the
conjecture. The more general rotating-center statement, without the
central-configuration condition, is given in \cref{prop:fixed-centers}.

Following \citet{RobertsMelanson2007}, we eliminate the velocity of a
constant-potential motion. Instead of expanding the resulting
polynomials and examining their Newton faces, we study a hypothetical
common algebraic curve on the complex distance cover. Two fixed
distances to distinct real centers determine only finitely many
positions. Consequently, some distance is nonconstant on the curve and
has a zero on its smooth projective completion. Complex isotropic
coordinates show that such a zero lies either over a finite position,
where exactly two distances vanish, or over infinity, where precisely
one distance vanishes. In both cases the velocity-compatibility equation
has a principal term that cannot be canceled.

For equilibrium finiteness, a critical curve would have constant
potential because the differential of \(\Omega\) vanishes on it.
The same endpoint analysis then produces a nonzero principal term of
one gradient component. Thus the critical curve is excluded directly,
without the Hessian calculation or the velocity-compatibility equation.
This also explains why no Morse assumption or preliminary isolation
theorem is needed. The dynamical argument requires the stronger
compatibility calculation, so the two main results have related but
different local obstructions.

The technical advantage of this formulation is uniform control of
exceptional configurations. The proof fixes the given masses and
centers throughout and proves the absence of positive-dimensional
components directly, rather than drawing conclusions only for regular
parameter values or for the already isolated part of the critical set.
An alleged critical curve is ruled out on its own
smooth projective model. At a finite position, distinct real centers
permit at most two zero distances; at infinity, just one zero distance
balances the quadratic term. Additional centers contribute only
lower-order terms. Consequently the same two local calculations work
for every finite number of centers, without symmetry assumptions or
an enumeration of parameter-dependent exceptional cases. For the
dynamical theorem, this curve argument replaces the explicit Newton-face
analysis used in the restricted three-body proof, while retaining the
velocity elimination of \citet{RobertsMelanson2007}.

The proof is organized as follows. \Cref{sec:model} derives the rotating
equations from a central configuration and records the Jacobi integral.
\Cref{sec:elimination} performs the velocity elimination.
\Cref{sec:algebraic} proves finiteness of the complex compatibility locus,
including all possible ends of a common curve. The restricted Saari
theorem and the real and complex equilibrium-finiteness theorems are proved in
\cref{sec:consequences}. The last section explains the essential scope
and the single-center exception.

\section{Central configurations and the restricted problem}\label{sec:model}

\subsection{From a central configuration to circular motion}

Fix an integer \(n\ge2\), and set \(N=n+1\). Let
\(m_1,\ldots,m_n>0\), and let \(a_1,\ldots,a_n\in\R^2\) be pairwise
distinct. We work in center-of-mass coordinates:
\begin{equation}\label{eq:cm}
 \sum_{j=1}^n m_ja_j=0.
\end{equation}
The configuration \(a=(a_1,\ldots,a_n)\) is central if there is a real
number \(\lambda\) such that
\begin{equation}\label{eq:cc}
 \sum_{\ell\ne j}m_\ell
       \frac{a_\ell-a_j}{|a_\ell-a_j|^3}
       =-\lambda a_j,\qquad j=1,\ldots,n.
\end{equation}
Define the primary potential and moment of inertia by
\begin{equation}\label{eq:primary-UI}
 U(a)=\sum_{i<j}\frac{m_im_j}{|a_i-a_j|},
 \qquad I(a)=\sum_{j=1}^n m_j|a_j|^2.
\end{equation}
Both are strictly positive. Taking the scalar product of the \(j\)-th
equation in \eqref{eq:cc} with \(m_ja_j\), and summing, gives
\begin{align*}
 -\lambda I(a)
 &=\sum_{i<j}\frac{m_im_j}{|a_i-a_j|^3}
   \bigl[a_i\cdot(a_j-a_i)+a_j\cdot(a_i-a_j)\bigr]\\
 &=-\sum_{i<j}\frac{m_im_j}{|a_i-a_j|}=-U(a).
\end{align*}
Thus
\begin{equation}\label{eq:lambda}
 \lambda=\frac{U(a)}{I(a)}>0.
\end{equation}
Choose \(\omega\in\R\setminus\{0\}\) with
\(\omega^2=\lambda\), and put
\begin{equation}\label{eq:R-J}
 R(\theta)=
 \begin{pmatrix}\cos\theta&-\sin\theta\\
                 \sin\theta& \cos\theta\end{pmatrix},
 \qquad
 J=\begin{pmatrix}0&-1\\1&0\end{pmatrix}.
\end{equation}
Then
\begin{equation}\label{eq:primaries}
 Q_j(t)=R(\omega t)a_j
\end{equation}
solves the Newtonian \(n\)-body equations. Indeed,
\(\ddot Q_j=-\omega^2R(\omega t)a_j\), while rotation of
\eqref{eq:cc} gives precisely the gravitational acceleration at
\(Q_j(t)\). Hence \eqref{eq:primaries} is the circular relative
equilibrium associated with \(a\).

This realization of the central configuration is part of the hypothesis.
A central configuration may also generate noncircular homographic
motions; those motions do not give the autonomous rotating system used
below. For background on central configurations and relative equilibria,
see \citet{MeyerHall1992}.

\subsection{The infinitesimal particle and the Jacobi integral}

Let \(Q(t)\in\R^2\) be the inertial position of a massless particle.
It has no effect on the primaries, and its acceleration is
\begin{equation}\label{eq:inertial}
 \ddot Q=\sum_{j=1}^n m_j
       \frac{Q_j(t)-Q}{|Q-Q_j(t)|^3}.
\end{equation}
Write \(Q(t)=R(\omega t)q(t)\). Since \(J^2=-\mathrm{Id}\),
\begin{equation}\label{eq:rotation-derivatives}
 \begin{split}
 \dot Q&=R(\omega t)(\dot q+\omega Jq),\\
 \ddot Q&=R(\omega t)(\ddot q+2\omega J\dot q-\omega^2q).
 \end{split}
\end{equation}
Substitution into \eqref{eq:inertial} gives
\begin{equation}\label{eq:rotating}
 \ddot q+2\omega J\dot q=\nabla\Omega(q),
\end{equation}
where the amended potential is
\begin{equation}\label{eq:Omega-real}
 \Omega(q)=\frac{\omega^2}{2}|q|^2+
         \sum_{j=1}^n\frac{m_j}{r_j},
 \qquad r_j=|q-a_j|.
\end{equation}
Its domain is
\(\cQ=\R^2\setminus\{a_1,\ldots,a_n\}\). A collision-free solution
will mean a classical solution \(q:\mathcal T\to\cQ\) on a connected
open interval \(\mathcal T\subset\R\).

Taking the scalar product of \eqref{eq:rotating} with \(\dot q\), and
using \(\dot q\cdot J\dot q=0\), shows that
\begin{equation}\label{eq:Jacobi}
 h=\frac12|\dot q|^2-\Omega(q)
\end{equation}
is constant. A different additive constant in \(\Omega\) changes only
the normalization of \(h\). An equilibrium is a point
\(q_*\in\cQ\) with \(\nabla\Omega(q_*)=0\).

Equation \eqref{eq:rotating} is the rotating system used in
\cref{thm:restricted}. The equivalence of constant amended potential
and constant rotating-frame speed follows from \eqref{eq:Jacobi}.
Stationarity plainly implies both conditions; the substantive content
of \cref{thm:restricted} is the converse. The speed in this statement
is the scalar speed in the rotating frame, not the velocity vector or
the inertial speed.

\section{Eliminating the velocity}\label{sec:elimination}

In this section the points \(a_j\) may be any pairwise distinct real
planar centers. They need not satisfy \eqref{eq:cm} or \eqref{eq:cc}.
We retain \(m_j>0\), \(n\ge2\), and \(\omega\ne0\), and consider
\eqref{eq:rotating} with \eqref{eq:Omega-real}.

For \(q=(x,y)\), write \(g=\nabla\Omega\) and define
\begin{equation}\label{eq:S-Lambda-real}
 \begin{split}
 S&=\Omega_x^2+\Omega_y^2,\\
 \Lambda&=\Omega_y^2\Omega_{xx}-2\Omega_x\Omega_y\Omega_{xy}
                         +\Omega_x^2\Omega_{yy}
          =(Jg)^T(\Hess\Omega)(Jg).
 \end{split}
\end{equation}
At real points \(S=|g|^2\). After complexification, \(S\) will always
denote the bilinear polynomial expression in
\eqref{eq:S-Lambda-real}, not a Hermitian norm.

\begin{lemma}[Velocity compatibility]\label{lem:velocity}
If a collision-free solution has \(\Omega(q(t))=c\), then
\(k=|\dot q(t)|^2\) is a nonnegative constant and
\begin{equation}\label{eq:P}
 P_k:=(S^2+k\Lambda)^2-4\omega^2kS^3=0
\end{equation}
at every point of that solution.
\end{lemma}

\begin{proof}
The first assertion follows from \eqref{eq:Jacobi}. If \(k=0\), then
\(q\) is constant; \eqref{eq:rotating} gives \(g=0\), and
\eqref{eq:P} is immediate. Suppose that \(k>0\).

At a point with \(g\ne0\), differentiating the constant-potential
condition gives
\(g\cdot\dot q=0\). In the plane the orthogonal complement of \(g\)
is spanned by \(Jg\), so locally
\begin{equation}\label{eq:velocity-tangent}
 \dot q=\sigma\sqrt{k}\frac{Jg}{\sqrt S},
 \qquad \sigma\in\{1,-1\}.
\end{equation}
A second derivative of the potential yields
\begin{align}
 0&=\frac{\dd^2}{\dd t^2}\Omega(q(t))
   =\dot q^T(\Hess\Omega)\dot q+g\cdot\ddot q\notag\\
  &=\frac{k\Lambda}{S}+g\cdot(g-2\omega J\dot q)
   =\frac{k\Lambda}{S}+S+2\sigma\omega\sqrt{kS}.
                                                        \label{eq:unsquared}
\end{align}
Multiplying by \(S\), rearranging, and squaring gives
\eqref{eq:P}. If \(g=0\) at a point, then \(S=\Lambda=0\) by
definition, and the same identity holds there. Thus no regularity
assumption on the potential level is needed.
\end{proof}

The elimination is the one introduced for the restricted three-body
problem by \citet{RobertsMelanson2007}, with the angular velocity left
unnormalized. Squaring enlarges the algebraic locus; we only use
\eqref{eq:P} as a necessary condition. It will suffice to prove
finiteness of this larger locus, without recovering the sign \(\sigma\)
or deciding which algebraic points are dynamically admissible.

\section{Finiteness on the complex distance cover}\label{sec:algebraic}

\subsection{Complex coordinates and rational derivatives}

Introduce independent complex coordinates
\begin{equation}\label{eq:zw}
 z=x+\ii y,\qquad w=x-\ii y,
\end{equation}
and fixed source coordinates
\begin{equation}\label{eq:alphabeta}
 \alpha_j=a_{jx}+\ii a_{jy},\qquad
 \beta_j=a_{jx}-\ii a_{jy},\qquad j=1,\ldots,n.
\end{equation}
Reality and distinctness of the centers imply that both families
\(\{\alpha_j\}\) and \(\{\beta_j\}\) are pairwise distinct. Also,
\begin{equation}\label{eq:real-distance}
 (\alpha_i-\alpha_j)(\beta_i-\beta_j)
 =|a_i-a_j|^2>0\qquad(i\ne j).
\end{equation}
The complex distance cover is
\begin{equation}\label{eq:cover}
 \cX=\left\{(z,w,r_1,\ldots,r_n):
 r_j^2=(z-\alpha_j)(w-\beta_j),\quad
 D:=\prod_{j=1}^n r_j\ne0\right\}.
\end{equation}
On it the amended potential is the rational function
\begin{equation}\label{eq:Omega-complex}
 \Omega=\frac{\omega^2}{2}zw+\sum_{j=1}^n\frac{m_j}{r_j}.
\end{equation}
The physical sheet has \(w=\overline z\) and \(r_j>0\). We do not
impose these conditions on the complex algebraic locus.

The condition \(D\ne0\) allows the coordinate derivatives to lift to
the cover. Indeed, the Jacobian of the distance relations with respect
to \((r_1,\ldots,r_n)\) is
\(\operatorname{diag}(2r_1,\ldots,2r_n)\), which is invertible on
\(\cX\). Thus \(\cX\) is a smooth complex surface and its projection
to the \((z,w)\)-plane is \'etale; locally \(z,w\) are holomorphic
coordinates and each \(r_j\) is a holomorphic function of them.
Differentiating the defining equations gives
\begin{equation}\label{eq:radius-derivatives}
 \partial_z r_j=\frac{w-\beta_j}{2r_j},
 \qquad
 \partial_w r_j=\frac{z-\alpha_j}{2r_j}.
\end{equation}
Consequently, we have
\begin{align}
 \Omega_z&=\frac{\omega^2}{2}w
       -\frac12\sum_{j=1}^n\frac{m_j(w-\beta_j)}{r_j^3},
                                                        \label{eq:Oz}\\
 \Omega_w&=\frac{\omega^2}{2}z
       -\frac12\sum_{j=1}^n\frac{m_j(z-\alpha_j)}{r_j^3},
                                                        \label{eq:Ow}
\end{align}
and
\begin{align}
 \Omega_{zz}&=\frac34\sum_{j=1}^n
                   \frac{m_j(w-\beta_j)^2}{r_j^5},
 &\Omega_{ww}&=\frac34\sum_{j=1}^n
                   \frac{m_j(z-\alpha_j)^2}{r_j^5},
                                                        \label{eq:pure-H}\\
 \Omega_{zw}&=\frac{\omega^2}{2}
                 +\frac14\sum_{j=1}^n\frac{m_j}{r_j^3}.
                                                        \label{eq:mixed-H}
\end{align}
For instance,
\[
 \partial_w\!\left(-\frac{m_j(w-\beta_j)}{2r_j^3}\right)
 =-\frac{m_j}{2r_j^3}
  +\frac{3m_j(z-\alpha_j)(w-\beta_j)}{4r_j^5}
 =\frac{m_j}{4r_j^3},
\]
which explains the sign of the source term in \eqref{eq:mixed-H}.

Since
\(\partial_x=\partial_z+\partial_w\) and
\(\partial_y=\ii(\partial_z-\partial_w)\),
\eqref{eq:S-Lambda-real} becomes
\begin{align}
 S&=4\Omega_z\Omega_w,                              \label{eq:S-complex}\\
 \Lambda&=-4\bigl(\Omega_w^2\Omega_{zz}
        -2\Omega_z\Omega_w\Omega_{zw}
        +\Omega_z^2\Omega_{ww}\bigr).               \label{eq:L-complex}
\end{align}
All derivatives in these formulas are computed on the distance cover
before restriction to a curve. They are not obtained by differentiating
a constant restriction \(\Omega|_\Gamma\).

\subsection{The affine compatibility locus}

Fix \(c,k\in\C\). Define \(\cZ_{c,k}\) by the distance relations,
the potential condition, and the velocity-compatibility condition:
\begin{equation}\label{eq:Z}
 \cZ_{c,k}=\{p\in\cX:\ \Omega(p)=c,\ P_k(p)=0\}.
\end{equation}
This is an affine algebraic set. To make the assertion explicit, adjoin
one coordinate \(s\) and use the following polynomial equations in
\(n+3\) variables:
\begin{equation}\label{eq:polynomial-system}
 \begin{gathered}
 r_j^2-(z-\alpha_j)(w-\beta_j)=0,
                    \qquad j=1,\ldots,n,\\
 sD-1=0,\\
 \left(\frac{\omega^2}{2}zw-c\right)D
       +\sum_{j=1}^n m_j\prod_{\ell\ne j}r_\ell=0,\\
 D^{24}P_k=0.
 \end{gathered}
\end{equation}
The last expression is a polynomial. Indeed, \(D^3\) clears the
denominators of \(\Omega_z,\Omega_w\), and \(D^5\) clears those of
the second derivatives. Therefore
\begin{equation}\label{eq:denominators}
 D^6S,\qquad D^{11}\Lambda,\qquad
 D^{12}(S^2+k\Lambda),\qquad D^{24}P_k
\end{equation}
are polynomials. These powers are sufficient for every finite \(n\);
their sharpness is immaterial. The equation \(sD=1\) excludes all
components introduced by vanishing denominators and determines \(s\)
uniquely. We identify \eqref{eq:Z} with this affine realization.

\begin{theorem}[Algebraic finiteness]\label{thm:finiteness}
Let \(n\ge2\), let \(a_1,\ldots,a_n\in\R^2\) be pairwise distinct,
let \(m_1,\ldots,m_n>0\), and let \(\omega\ne0\). For every
\(c,k\in\C\), the complex algebraic set \(\cZ_{c,k}\) is finite.
\end{theorem}

The number of equations and variables in
\eqref{eq:polynomial-system} alone does not prove finiteness. We shall
exclude every possible common curve, including curves that might occur
at exceptional values of the parameters.

\subsection{A nonconstant distance has a zero}

We use standard facts about affine algebraic sets and the smooth
projective models of algebraic curves; see
\citet[Chap.~I, especially \S6]{Hartshorne1977}.

\begin{lemma}[Finite constant-distance fibers]\label{lem:fibers}
For two distinct real centers, fixing their two distance coordinates on
the complex distance cover permits at most two values of \((z,w)\).
Consequently, every algebraic curve in \(\cX\) has at least one
nonconstant distance coordinate.
\end{lemma}

\begin{proof}
Fix \(r_1,r_2\). Subtracting their squared-distance equations gives
\begin{equation}\label{eq:radical-axis}
 (\beta_2-\beta_1)z+(\alpha_2-\alpha_1)w
  =r_1^2-r_2^2-\alpha_1\beta_1+\alpha_2\beta_2.
\end{equation}
Both coefficients on the left are nonzero by
\eqref{eq:real-distance}. Solving for \(w\) and substituting in
\(r_1^2=(z-\alpha_1)(w-\beta_1)\) gives a quadratic equation in
\(z\) whose leading coefficient is
\[
 -\frac{\beta_2-\beta_1}{\alpha_2-\alpha_1}\ne0.
\]
There are therefore at most two positions. If all distances were
constant on an irreducible curve, every remaining coordinate, including
\(s=1/D\), would also take finitely many values. That is impossible
for an algebraic curve.
\end{proof}

\begin{lemma}[Vanishing distance on a compact model]\label{lem:zero}
If \(\cZ_{c,k}\) is infinite, it contains an irreducible affine
algebraic curve \(\Gamma\). On the smooth projective model
\(\widehat\Gamma\) of this curve, some distance \(r_i\) is nonconstant
and has a zero.
\end{lemma}

\begin{proof}
An infinite affine algebraic set over \(\C\) has a positive-dimensional
irreducible component. Such a component contains an irreducible affine
curve, obtained, if necessary, by taking hyperplane sections. Apply this
to \eqref{eq:polynomial-system}. By \cref{lem:fibers}, some \(r_i\)
is nonconstant on the resulting curve \(\Gamma\).

The coordinate functions of \(\Gamma\) are rational functions on its
smooth projective model and hence meromorphic on the compact Riemann
surface \(\widehat\Gamma\). If the nonconstant function \(r_i\) had no
zero there, then \(1/r_i\) would have no pole and would be holomorphic
on a compact Riemann surface. It would be constant, a contradiction.
Thus \(r_i\) has a zero on \(\widehat\Gamma\).
\end{proof}

Fix such a zero \(p\), and write \(\varepsilon=r_i\). Choose a true
local uniformizer \(t\) at \(p\). Then
\begin{equation}\label{eq:uniformizer}
 \varepsilon=t^e u(t),\qquad e\ge1,\qquad u(0)\ne0.
\end{equation}
The use of \(\varepsilon\) in asymptotic notation does not assume
\(e=1\). All estimates below are statements about orders in \(t\).
For example, \(f=O(\varepsilon^b)\) means
\(\ord_p f\ge be\), and \(f\sim A\varepsilon^b\) means that the
quotient has the indicated nonzero limit along a local Puiseux branch.
Every coordinate is meromorphic at \(p\), so \(z,w\) are either finite
or have poles there.

The two following lemmas exhaust all possibilities. Their essential
feature is that any additional centers contribute only lower-order
terms.

\subsection{A zero over a finite position}

\begin{lemma}\label{lem:finite-end}
The zero \(p\) in \cref{lem:zero} cannot lie over a finite value of both
\(z\) and \(w\).
\end{lemma}

\begin{proof}
Let \(z(p)=z_0\) and \(w(p)=w_0\) be finite. A vanishing distance
\(r_j\) requires \(z_0=\alpha_j\) or \(w_0=\beta_j\). Since the
\(\alpha_j\) are distinct and the \(\beta_j\) are distinct, at most
two distances can vanish at \(p\), independently of \(n\).

Exactly one cannot vanish: its term \(m_j/r_j\) would be the only pole
of \(\Omega\), while \(\Omega=c\) on \(\Gamma\). Thus exactly two
distances vanish. After an interchange of indices or of \(z,w\), we
have
\begin{equation}\label{eq:finite-position}
 z\to\alpha_i,\qquad w\to\beta_j,\qquad
 r_i=\varepsilon\to0,\qquad r_j\to0,\qquad i\ne j.
\end{equation}
Every \(r_\ell\) with \(\ell\notin\{i,j\}\) has a finite nonzero
limit. The remaining terms of the potential are therefore holomorphic
at \(p\). Write the potential identity as
\begin{equation}\label{eq:finite-cancellation}
 \frac{m_i}{\varepsilon}+\frac{m_j}{r_j}=H(t),
\end{equation}
where \(H\) is holomorphic. Solving gives
\begin{equation}\label{eq:rho}
 r_j=-\frac{m_j\varepsilon}{m_i-\varepsilon H(t)}
     =\rho\varepsilon+O(\varepsilon^2),
 \qquad \rho=-\frac{m_j}{m_i}\ne0.
\end{equation}
In particular, \(\ord_p r_j=e\). The distance relations also imply
\begin{equation}\label{eq:finite-shifts}
 z-\alpha_i=\frac{\varepsilon^2}{w-\beta_i}
          =O(\varepsilon^2),
 \qquad
 w-\beta_j=\frac{r_j^2}{z-\alpha_j}
          =O(\varepsilon^2).
\end{equation}
Both denominators tend to nonzero limits, so the two orders are
exactly \(2e\).

In \(\Omega_z\), the \(i\)-th term has a pole of order \(3e\) with
nonzero principal coefficient, whereas the \(j\)-th term is only
\(O(\varepsilon^{-1})\) by \eqref{eq:finite-shifts}. All other source
terms are bounded. The roles of \(i,j\) are reversed in \(\Omega_w\).
It follows that
\begin{equation}\label{eq:finite-gradient}
 \Omega_z\sim-\frac{m_i(\beta_j-\beta_i)}{2\varepsilon^3},
 \qquad
 \Omega_w\sim-\frac{m_j(\alpha_i-\alpha_j)}{2r_j^3}.
\end{equation}
Using \eqref{eq:real-distance}, \eqref{eq:S-complex}, and
\eqref{eq:rho}, we obtain
\begin{equation}\label{eq:finite-S}
 S\sim-\frac{m_im_j|a_i-a_j|^2}{\varepsilon^3r_j^3}
   \sim A_{ij}\varepsilon^{-6},
 \qquad
 A_{ij}=\frac{m_i^4|a_i-a_j|^2}{m_j^2}\ne0.
\end{equation}

Every second derivative in \eqref{eq:pure-H}--\eqref{eq:mixed-H}
is \(O(\varepsilon^{-5})\); the first derivatives are
\(O(\varepsilon^{-3})\). Therefore \eqref{eq:L-complex} gives
\begin{equation}\label{eq:finite-L}
 \Lambda=O(\varepsilon^{-11}).
\end{equation}
In valuation language,
\(\ord_p S=-6e\) and \(\ord_p\Lambda\ge-11e\). Hence
\begin{equation}\label{eq:finite-ratios}
 \frac{\Lambda}{S^2}=O(\varepsilon)\to0,
 \qquad \frac1S\to0.
\end{equation}
Since \(P_k=0\) as a rational identity on \(\Gamma\), and
\eqref{eq:finite-S} permits division by \(S\) near \(p\), we arrive at
\begin{equation}\label{eq:contradiction}
 0=\frac{P_k}{S^4}
   =\left(1+\frac{k\Lambda}{S^2}\right)^2
          -\frac{4\omega^2k}{S}
   \longrightarrow1.
\end{equation}
This contradiction excludes the finite case.
\end{proof}

\subsection{A zero over complex infinity}

\begin{lemma}\label{lem:infinite-end}
The zero \(p\) in \cref{lem:zero} cannot be a pole of \(z\) or of
\(w\).
\end{lemma}

\begin{proof}
The identity
\begin{equation}\label{eq:small-radius}
 \varepsilon^2=(z-\alpha_i)(w-\beta_i)
\end{equation}
prevents both \(z\) and \(w\) from having poles at \(p\). If one has
a pole, the other must approach the corresponding source coordinate.
After interchanging \(z,w\), we may suppose that
\begin{equation}\label{eq:infinite-position}
 z\to\alpha_i,\qquad w\to\infty.
\end{equation}
For each \(j\ne i\),
\begin{equation}\label{eq:large-radii}
 r_j^2=(z-\alpha_j)(w-\beta_j)
        \sim(\alpha_i-\alpha_j)w.
\end{equation}
Thus all other radii have poles and \(1/r_j\to0\).

There is a special case when the \(i\)-th center is the center of
rotation. If \(\alpha_i=0\), reality implies \(\beta_i=0\), and
\eqref{eq:small-radius} gives \(zw=\varepsilon^2\to0\). Then the
quadratic term and all source terms other than \(m_i/\varepsilon\)
tend to zero. They cannot cancel its pole in \(\Omega=c\). Hence
\begin{equation}\label{eq:alpha-nonzero}
 \alpha_i\ne0.
\end{equation}

Using \(z=\alpha_i+\varepsilon^2/(w-\beta_i)\), the potential identity
takes the exact form
\begin{equation}\label{eq:infinite-balance}
 \frac{\omega^2\alpha_i}{2}w+\frac{m_i}{\varepsilon}
 +\frac{\omega^2}{2}\frac{\varepsilon^2w}{w-\beta_i}
 +\sum_{j\ne i}\frac{m_j}{r_j}=c.
\end{equation}
The third term is \(O(\varepsilon^2)\), and the last sum tends to
zero by \eqref{eq:large-radii}. Consequently, the first two terms have
equal pole orders and opposite principal coefficients. Set
\begin{equation}\label{eq:K}
 K_i=-\frac{2m_i}{\omega^2\alpha_i}\ne0.
\end{equation}
Then
\begin{equation}\label{eq:infinite-wz}
 w=\frac{K_i}{\varepsilon}+O(1),
 \qquad
 z-\alpha_i=\frac{\varepsilon^3}{K_i}+O(\varepsilon^4).
\end{equation}
In particular, \(\ord_p w=-e\) and
\(\ord_p(z-\alpha_i)=3e\). Equation \eqref{eq:large-radii} further
gives
\begin{equation}\label{eq:radius-orders}
 \ord_p r_j=-\frac e2\qquad(j\ne i).
\end{equation}
Since these radii are meromorphic, their orders are integers. Thus
\(e\) is even. This explains the half-integer powers of
\(\varepsilon\) in the estimates below.

We next compute the gradient, retaining the interaction between the
singular source and the centrifugal term. For \(\Omega_z\), the
singular source dominates:
\begin{equation}\label{eq:infinite-Oz}
 \Omega_z\sim-\frac{m_iK_i}{2}\varepsilon^{-4}
            =\frac{m_i^2}{\omega^2\alpha_i}\varepsilon^{-4}.
\end{equation}
For \(\Omega_w\), both the centrifugal term and the \(i\)-th source
have nonzero finite limits. By \eqref{eq:infinite-wz},
\begin{equation}\label{eq:infinite-Ow}
 \Omega_w\longrightarrow
       \frac{\omega^2\alpha_i}{2}-\frac{m_i}{2K_i}
       =\frac{3\omega^2\alpha_i}{4}\ne0.
\end{equation}
For \(j\ne i\), the source contributions to \(\Omega_z\) and
\(\Omega_w\) are respectively
\(O(\varepsilon^{1/2})\) and \(O(\varepsilon^{3/2})\): their
numerators have orders \(\varepsilon^{-1}\) and \(1\), whereas
\(r_j^{-3}=O(\varepsilon^{3/2})\). There are only finitely many
such terms. Thus
\begin{equation}\label{eq:infinite-S}
 S=3m_i^2\varepsilon^{-4}+O(\varepsilon^{-3}).
\end{equation}

The second derivatives satisfy
\begin{equation}\label{eq:infinite-H}
 \Omega_{zz}=O(\varepsilon^{-7}),\qquad
 \Omega_{zw}=O(\varepsilon^{-3}),\qquad
 \Omega_{ww}=O(\varepsilon).
\end{equation}
Indeed, the singular-source numerators in the two pure derivatives have
orders \(\varepsilon^{-2}\) and \(\varepsilon^6\), respectively,
and their denominator is \(\varepsilon^5\). The other centers
contribute respectively
\(O(\varepsilon^{1/2})\), \(O(\varepsilon^{3/2})\), and
\(O(\varepsilon^{5/2})\) to
\(\Omega_{zz},\Omega_{zw},\Omega_{ww}\). All these contributions
are weaker than the orders displayed in \eqref{eq:infinite-H}.

In \eqref{eq:L-complex}, the three products have orders at least
\(-7e\): they are the products of quantities of orders
\((0,-7e)\), \((-4e,0,-3e)\), and \((-8e,e)\), respectively.
Therefore
\begin{equation}\label{eq:infinite-L}
 \Lambda=O(\varepsilon^{-7}),\qquad
 \ord_p S=-4e,\qquad \ord_p\Lambda\ge-7e.
\end{equation}
Together with \eqref{eq:infinite-S}, this yields
\begin{equation}\label{eq:infinite-ratios}
 \frac{\Lambda}{S^2}=O(\varepsilon)\to0,
 \qquad \frac1S\to0.
\end{equation}
The contradiction \eqref{eq:contradiction} follows again. This excludes
every infinite zero of a distance function.
\end{proof}

\subsection{Completion of the finiteness proof}

\begin{proof}[Proof of \cref{thm:finiteness}]
If \(\cZ_{c,k}\) were infinite, \cref{lem:zero} would give an
irreducible curve and a zero \(p\) of a nonconstant distance on its
smooth projective model. Since the coordinates are meromorphic, either
both \(z,w\) are finite at \(p\), or at least one has a pole.
\Cref{lem:finite-end,lem:infinite-end} exclude these possibilities.
Thus there is no algebraic curve in \(\cZ_{c,k}\). Its dimension is
at most zero, and it is finite, with the empty set allowed.
\end{proof}

The proof treats all \(c,k\in\C\), not merely regular real levels or
generic parameter values. No estimate for the number of points is
required. The local contradictions take place on the same curve that
would satisfy both algebraic conditions, so no irreducibility claim for
the entire potential level is needed.

\section{The restricted theorem and equilibrium finiteness}\label{sec:consequences}

\subsection{Constant amended potential forces stationarity}

\begin{proposition}[Fixed centers in a rotating frame]\label{prop:fixed-centers}
Let \(n\ge2\), let \(a_1,\ldots,a_n\in\R^2\) be pairwise distinct,
and let \(m_j>0\), \(\omega\ne0\). Every collision-free solution of
\eqref{eq:rotating} with constant \(\Omega\) is stationary.
\end{proposition}

\begin{proof}
Let \(\Omega(q(t))=c\). The Jacobi integral gives
\(k=|\dot q(t)|^2\), constant. If \(k=0\), then \(q\) is constant
and \eqref{eq:rotating} gives \(\nabla\Omega(q)=0\). Suppose that
\(k>0\). By \cref{lem:velocity}, the continuous lifted trajectory
\[
 t\longmapsto
 \bigl(x(t)+\ii y(t),\ x(t)-\ii y(t),\
       r_1(t),\ldots,r_n(t),\ D(t)^{-1}\bigr)
\]
lies in \(\cZ_{c,k}\). This set is finite by \cref{thm:finiteness}.
The image of a connected interval in a finite set is a point. Hence
\(q\) is constant, contrary to \(k>0\). Thus only the stationary
case is possible.
\end{proof}

\begin{proof}[Proof of \cref{thm:restricted}]
The central-configuration equations and \eqref{eq:lambda} produce the
circular primary motion \eqref{eq:primaries}. The transformation
\eqref{eq:rotation-derivatives} gives precisely the system in
\cref{prop:fixed-centers}. That proposition proves the implication
from constant amended potential to equilibrium. The equivalence with
constant rotating-frame speed follows from \eqref{eq:Jacobi}. Finally,
\(Q(t)=R(\omega t)q_*\) is the inertial realization of an equilibrium.
\end{proof}

\subsection{Complex critical-locus finiteness and the planar main theorem}

We now prove the stronger complex statement underlying
\cref{thm:equilibria}. The proof uses the same meromorphic distance
functions as the dynamical argument, but only their first derivatives
are needed. In particular, the endpoint balances through
\eqref{eq:finite-gradient} and \eqref{eq:infinite-Ow} use solely the
distance relations and \(\Omega=c\), not \(P_k=0\).

\begin{theorem}[Complex critical-locus finiteness]\label{thm:complex-equilibria}
Let \(n\ge2\), let \(a_1,\ldots,a_n\in\R^2\) be pairwise distinct,
let \(m_1,\ldots,m_n>0\), and let
\(\omega\in\R\setminus\{0\}\). On the complex distance cover
\(\cX\) of \eqref{eq:cover}, the algebraic set
\begin{equation}\label{eq:critical-locus}
 \Crit_{\cX}(\Omega)=
 \{p\in\cX:\ \Omega_z(p)=\Omega_w(p)=0\}
\end{equation}
is finite. The conclusion includes every sheet of the cover, not only
the physical sheet \(w=\overline z\), \(r_j>0\).
\end{theorem}

\begin{proof}
Adjoin \(sD=1\) as before. In addition to the distance relations,
impose \(2D^3\Omega_z=2D^3\Omega_w=0\). These are polynomial
equations; for example,
\begin{equation}\label{eq:critical-polynomial}
 2D^3\Omega_z=\omega^2wD^3-
  \sum_{j=1}^n m_j(w-\beta_j)\prod_{\ell\ne j}r_\ell^3.
\end{equation}
The corresponding expression for \(2D^3\Omega_w\) is obtained by
interchanging \(z,\alpha_j\) with \(w,\beta_j\). Because \(sD=1\),
clearing denominators creates no collision components and does not
discard any point of \eqref{eq:critical-locus}. Hence this is an affine
algebraic realization of the critical locus.

If it were infinite, it would contain an irreducible affine curve
\(\Gamma\). On the smooth distance cover the lifted derivative
identities \eqref{eq:radius-derivatives} give the differential identity
\begin{equation}\label{eq:critical-differential}
 \dd\Omega=\Omega_z\dd z+\Omega_w\dd w.
\end{equation}
Pull this identity back to the smooth projective model
\(\widehat\Gamma\). Both coefficients vanish on the dense affine
part of the curve, so the resulting meromorphic differential is
identically zero. This remains valid if the affine curve is singular,
since the pullback is taken to its normalization. To see the constancy
explicitly, a local Laurent expansion \(\Omega=\sum_\ell b_\ell t^\ell\)
has derivative \(\sum_\ell\ell b_\ell t^{\ell-1}\dd t=0\); in
characteristic zero all coefficients with \(\ell\ne0\) vanish.
Thus \(\Omega=c\) on the connected compact model, for some \(c\in\C\).

By \cref{lem:fibers}, at least one distance \(r_i\) is nonconstant
on \(\Gamma\). As in the proof of \cref{lem:zero}, it is meromorphic
on \(\widehat\Gamma\) and has a zero \(p\). Write
\(\varepsilon=r_i=t^eu(t)\), with \(e\ge1\), \(u(0)\ne0\), in a
true local uniformizer. There are two possibilities.

If \(z,w\) are finite at \(p\), constancy of the potential forces
exactly two zero distances. After interchanging indices or coordinates,
they are \(r_i,r_j\), with \(z\to\alpha_i\),
\(w\to\beta_j\), and \(i\ne j\). The cancellation in
\eqref{eq:finite-cancellation} gives
\(r_j=-(m_j/m_i)\varepsilon+O(\varepsilon^2)\), and
\eqref{eq:finite-gradient} yields
\[
 \Omega_z\sim
 -\frac{m_i(\beta_j-\beta_i)}{2}\varepsilon^{-3}.
\]
Its coefficient is nonzero because the real centers are distinct.
This contradicts \(\Omega_z|_\Gamma=0\).

If one coordinate has a pole, the distance relation prevents both
from having poles. Interchanging \(z,w\) if necessary gives
\(z\to\alpha_i\), \(w\to\infty\). The case \(\alpha_i=0\)
is already incompatible with \(\Omega=c\), as shown before
\eqref{eq:alpha-nonzero}. Otherwise the potential balance
\eqref{eq:infinite-balance} gives \eqref{eq:infinite-wz}, and hence
\[
 \Omega_w\longrightarrow\frac{3\omega^2\alpha_i}{4}\ne0
\]
by \eqref{eq:infinite-Ow}. This contradicts
\(\Omega_w|_\Gamma=0\). Thus every possible zero of a nonconstant
distance is excluded. There is no critical curve; the affine critical
locus has dimension at most zero and is finite.
\end{proof}

\begin{proof}[Proof of \cref{thm:equilibria}]
For each real point \(q=(x,y)\) outside the centers, choose its unique
physical lift
\[
 z=x+\ii y,\quad w=x-\ii y,\quad r_j=|q-a_j|>0.
\]
The identities \(\Omega_x=\Omega_z+\Omega_w\) and
\(\Omega_y=\ii(\Omega_z-\Omega_w)\) show that a real critical point
lifts to \(\Crit_{\cX}(\Omega)\). The lift is injective, and the
latter set is finite by \cref{thm:complex-equilibria}. Hence the
real critical set is finite. This argument imposes no condition on
the Hessian and includes all degenerate equilibria.
\end{proof}

The complex theorem can alternatively be deduced from
\cref{thm:finiteness}: on a critical curve the preceding differential
argument gives \(\Omega=c\), while \(S=\Lambda=0\) implies
\(P_0=0\). Such a curve would be contained in the finite set
\(\cZ_{c,0}\). The direct proof above displays the first-derivative
obstruction and separates equilibrium finiteness from velocity
elimination.

For comparison with the known restricted result, suppose the centers
do satisfy \eqref{eq:cc} with \(\lambda=\omega^2\). At a real
critical point \(q\),
\[
 \sum_{j=1}^n m_j\frac{a_j-q}{|a_j-q|^3}=-\omega^2q.
\]
Adding a particle of mass zero at \(q\) leaves the primary equations
and center of mass unchanged, and the displayed equation is precisely
its central-configuration equation. Thus real finiteness in this
special case follows already from
\citet[Theorem~3.1]{HamptonJensen2015}. The arbitrary-center statement
of \cref{thm:equilibria} does not require these primary equations.
The new conclusion is unconditional finiteness in that larger class,
and, more strongly, finiteness of the full complex critical locus.

\section{Scope, sharpness, and the role of the assumptions}\label{sec:scope}

\subsection{Why at least two distinct centers are needed}

The condition \(n\ge2\) is essential to the fixed-center result.
For one source of mass \(m>0\) at the origin,
\begin{equation}\label{eq:one-center-potential}
 \Omega(q)=\frac{\omega^2}{2}|q|^2+\frac{m}{|q|}.
\end{equation}
Choose a radius \(\rho>0\), put
\(\sigma=\sqrt{m/\rho^3}\), and consider
\begin{equation}\label{eq:one-center-orbit}
 q(t)=\rho\bigl(\cos((\sigma-\omega)t),
                   \sin((\sigma-\omega)t)\bigr).
\end{equation}
It is the rotating-frame expression of a circular Kepler orbit.
Directly,
\[
 \ddot q+2\omega J\dot q
 =\bigl(\omega^2-\sigma^2\bigr)q
 =\nabla\Omega(q).
\]
The potential is constant along this orbit, but the orbit is not
stationary unless \(\sigma=\omega\). Moreover,
\(\nabla\Omega=0\) on the entire circle
\(|q|=(m/\omega^2)^{1/3}\), so equilibrium finiteness fails too.

This is precisely where \cref{lem:fibers} would fail: a single fixed
distance permits a circle of positions. Coincident source locations
must likewise be combined before applying the argument; one effective
center cannot be treated as several distinct centers.

\subsection{Planarity and circular realization}

Two uses of planarity are fundamental. The velocity orthogonal to a
nonzero gradient is fixed up to sign and magnitude, which makes
\eqref{eq:velocity-tangent} possible. Also, a squared planar distance
factors as \((z-\alpha_j)(w-\beta_j)\). Distinct real centers therefore
allow at most two simultaneous zero distances over a finite complex
position. Neither statement has the same form for spatial motion.
The present proof makes no assertion about a spatial restricted
\(N\)-body analogue.

Likewise, a central configuration by itself is a shape, not a choice of
its evolution. The circular realization supplies one constant angular
velocity and an autonomous amended potential. An elliptic or other
noncircular homographic realization leads instead to time-dependent
rotating or pulsating equations. The Jacobi integral
\eqref{eq:Jacobi} and the compatibility equation \eqref{eq:P} cannot
be carried over to that setting without a separate derivation.





\appendix
\section{Principal coefficients and an alternative tensor identity}
\label{app:coefficients}

The orders used in the proof are sufficient for the contradictions.
We record the principal coefficients and a coordinate-free identity as
independent checks on those estimates.

\subsection{The two nonzero principal parts}

At a finite end, \eqref{eq:finite-S} and \eqref{eq:finite-L} imply
\begin{equation}\label{eq:finite-P-principal}
 P_k\sim A_{ij}^4\varepsilon^{-24}
       =\frac{m_i^{16}|a_i-a_j|^8}{m_j^8}\varepsilon^{-24}.
\end{equation}
Indeed, \(S^2\) has order \(-12e\), whereas \(k\Lambda\) has order
at least \(-11e\), and \(S^3\) has order \(-18e\). Thus the
\(S^4\) term is the unique principal contribution to \(P_k\).

At an infinite end, write \(\nu=\omega^2\), \(m=m_i\),
\(\alpha=\alpha_i\). From \eqref{eq:infinite-wz} and the derivative
formulas,
\begin{align*}
 \Omega_z&\sim\frac{m^2}{\nu\alpha}\varepsilon^{-4},
 &\Omega_w&\sim\frac{3\nu\alpha}{4},\\
 \Omega_{zz}&\sim\frac{3m^3}{\nu^2\alpha^2}\varepsilon^{-7},
 &\Omega_{zw}&\sim\frac m4\varepsilon^{-3},
 &\Omega_{ww}&\sim\frac{3\nu^2\alpha^2}{16m}\varepsilon.
\end{align*}
The second line follows by substituting
\(w\sim-2m/(\nu\alpha\varepsilon)\) and
\(z-\alpha\sim-\nu\alpha\varepsilon^3/(2m)\) into the singular
source terms. The other sources have strictly higher orders. Equation
\eqref{eq:L-complex} then gives
\begin{equation}\label{eq:infinite-L-principal}
 \Lambda\sim
 -4\left(\frac{27}{16}-\frac{3}{8}+\frac{3}{16}\right)
 m^3\varepsilon^{-7}
 =-6m^3\varepsilon^{-7}.
\end{equation}
Combining this with \eqref{eq:infinite-S} yields
\begin{equation}\label{eq:infinite-P-principal}
 P_k\sim81m^8\varepsilon^{-16}.
\end{equation}
The coefficients in \eqref{eq:finite-P-principal} and
\eqref{eq:infinite-P-principal} are independent of \(c\), \(k\), and
every source not involved in the corresponding singular balance.

\subsection{A cofactor formula}

After complexification use the bilinear dot product, put
\(v_j=q-a_j\), and recall that \(v_j\cdot v_j=r_j^2\). Direct
differentiation gives
\begin{equation}\label{eq:real-tensor}
 g=\omega^2q-\sum_j\frac{m_jv_j}{r_j^3},
 \qquad
 H:=\Hess\Omega
   =\omega^2\mathrm{Id}
     +\sum_jm_j\left(-\frac{\mathrm{Id}}{r_j^3}
                   +\frac{3v_jv_j^T}{r_j^5}\right).
\end{equation}
For a symmetric two-by-two matrix,
\(J^THJ=\cof(H)=(\operatorname{tr}H)\mathrm{Id}-H\).
Consequently,
\begin{equation}\label{eq:cofactor}
 \Lambda=g^T\cof(H)g
  =\omega^2S+
   \sum_jm_j\left(\frac{2S}{r_j^3}
                -\frac{3(g\cdot v_j)^2}{r_j^5}\right).
\end{equation}
This agrees with \eqref{eq:L-complex}. At the infinite end one also
has
\[
 g\cdot v_i=-\frac{2m_i}{\varepsilon}+O(1).
\]
The \(i\)-th summand in \eqref{eq:cofactor} therefore has principal
part \((6-12)m_i^3\varepsilon^{-7}\), recovering
\eqref{eq:infinite-L-principal} without expanding the complex formula.

\section*{Acknowledgements}

\end{document}